\documentclass[aps,prl,nobibnotes,twocolumn,superscriptaddress,english]{revtex4-2}
\usepackage[utf8]{inputenc}
\usepackage{amsmath}
\usepackage{amssymb}
\usepackage{graphicx}
\usepackage{xcolor}
\usepackage{siunitx}
\usepackage{hyperref}

\newcommand{\Tc}{$T_\text{c}$}
\newcommand{\pc}{$p_\text{c}$}
\newcommand{\BSH}{BaSiH$_8$}
\newcommand{\SSH}{SrSiH$_8$}
\newcommand{\LBH}{LaBH$_8$}
\newcommand{\XYH}{$XY$H$_8$}
\newcommand{\fmm}{$Fm\bar{3}m$}

\makeatletter
\def\@fnsymbol#1{\ensuremath{\ifcase#1\or \dagger\or *\or \ddagger\or
   \mathsection\or \mathparagraph\or \|\or **\or \dagger\dagger
   \or \ddagger\ddagger \else\@ctrerr\fi}}
\makeatother

\begin{document}

\author{Roman Lucrezi}
\affiliation{Institute of Theoretical and Computational Physics, Graz University of Technology, NAWI Graz, 8010 Graz, Austria}
\author{Simone Di Cataldo}
\affiliation{Institute of Theoretical and Computational Physics, Graz University of Technology, NAWI Graz, 8010 Graz, Austria}
\affiliation{Dipartimento di Fisica, Sapienza Universit\`a di Roma, 00185 Rome, Italy} 
\author{Wolfgang von der Linden}
\affiliation{Institute of Theoretical and Computational Physics, Graz University of Technology, NAWI Graz, 8010 Graz, Austria}
\author{Lilia Boeri}\email{lilia.boeri@uniroma1.it}
\affiliation{Dipartimento di Fisica, Sapienza Universit\`a di Roma, 00185 Rome, Italy} 
\affiliation{Enrico Fermi Research Center, Via Panisperna 89 A, 00184 Rome, Italy}
\author{Christoph Heil} \email{christoph.heil@tugraz.at}
\affiliation{Institute of Theoretical and Computational Physics, Graz University of Technology, NAWI Graz, 8010 Graz, Austria}

\title{In-silico synthesis of lowest-pressure high-\texorpdfstring{\Tc}{Tc} ternary superhydrides}

\date{\today}
\begin{abstract}
We report the theoretical prediction of two high-performing hydride superconductors \BSH \ and \SSH. They are thermodynamically stable above pressures of 130 and \SI{174}{GPa}, respectively, and metastable below that. Employing anharmonic phonon calculations, we determine the minimum pressures of dynamical stability to be around \SI{3}{GPa} for \BSH \ and \SI{27}{GPa} for \SSH, and using the fully anisotropic Migdal-Eliashberg theory, we predict \Tc's around 71 and \SI{126}{K}, respectively. 
We also introduce a method to estimate the lowest pressure of synthesis,
based on the calculation of the enthalpy barriers protecting the \BSH \ \fmm \ structure from decomposition at various pressures. This {\em kinetic} pressure threshold is sensibly higher than the one based on {\em dynamic} stability, but gives a much more rigorous limit for synthesizability.
\end{abstract}

\maketitle

\section{Introduction}
The unexpected discovery of high-temperature superconductivity in high-pressure hydrides has revamped the 
hopes of ending the century-long quest for superconductivity
at ambient conditions~\cite{ashcroft1968metallic,ashcroft2004hydrogen,drozdov2015conventional,drozdov2019superconductivity,somayazulu2019evidence,flores2020perspective,sun2019route,semenok2020superconductivity,snider2020room,troyan2021anomalous,kong2019superconductivity,chen2021high,boeri_roadmap_2021}. In less than five years, superhydrides have established
higher and higher records for critical temperatures, starting
with SH$_3$ (\SI{203}{K})~\cite{drozdov2015conventional}, LaH$_{10}$ (\SI{265}{K})~\cite{somayazulu2019evidence, drozdov2019superconductivity}, and C-S-H (\SI{288}{K})~\cite{snider2020room}.

While it is very exciting to strive for new superconductors 
with even higher \Tc's attaining 
{\em hot superconductivity}~\cite{peng2017hydrogen, grockowiak2020hot, di2021x}, 
lowering the required stabilization pressures while
retaining \Tc \ above the boiling point of nitrogen (\SI{77}{K})
is of even greater importance~\cite{pickard2020superconducting,lv2020theory,di2021labh,shipley2021high,zhang2021design,di2020phase,di2021x}.
In fact, the discovery of a conventional ($s$-wave) superconductor
with these properties would open up a wide array of technological applications in key strategical sectors such as energy conservation, climate change, and medicine.

After exhausting the search space of binary hydrides, the focus of superhydride research is rapidly shifting to ternary hydrides, where the parameter space is much larger~\cite{flores2020perspective,pickard2020superconducting}.
We have recently predicted, for example, that a ternary hydride with \LBH \ composition could form an \fmm \ ternary sodalite-like clathrate (SLC) structure and remain stable down to a critical pressure $p_\text{c} \sim \SI{35}{GPa}$, with a $T_\text{c} = \SI{126}{K}$~\cite{di2021labh,di2021x}.
Our findings were later confirmed by independent studies on the La-B-H system~\cite{zhang2021design,liang2021prediction}. 
The predicted \pc \ of \LBH \ is a factor of four
lower than in binary hydrides, where \pc's are in
the Megabar range, but  still too high to envision any large-scale applications for this particular compound.  However, through
the identification of the \fmm \ \XYH \ structural template,
the discovery of \LBH \ paved the road to the study of a whole new family of potential high-\Tc \ ternary hydrides, where \pc \
and \Tc \ may be improved even further.

In this work, using first-principles methods for crystal structure prediction and superconductivity, we identify two high-\Tc \ alkaline-earth/silicon superhydrides, \BSH \ and \SSH.
We predict that both compounds will spontaneously form in the \fmm \ \XYH \ structure at high pressures (130 and \SI{174}{GPa}, respectively), and remain dynamically stable down to much lower pressures ($p_\text{c} = 3$ and $\SI{27}{GPa}$), with superconducting \Tc's of 71 and \SI{126}{K}, respectively. 

We also succeed in deriving a  rigorous limit for the stability of \BSH, calculating explicitly the energy barrier protecting the metastable \fmm \ structure from decomposition (\textit{kinetic} stability) as a function of pressure, using the Variable-Cell Nudged Elastic Band (VCNEB) method~\cite{oganov_2013_VCNEB}.
We find that,  indeed, \fmm \ \BSH \ should remain metastable
at pressures well below \SI{130}{GPa}. However, the {\em kinetic} critical pressure $p_\text{kin}$ determined by the energy barrier is significantly higher (\SI{30}{GPa}) than the value estimated from anharmonic lattice dynamics.
Similar discrepancies between dynamical and kinetic pressure of stability may explain the systematic underestimation of the predicted pressures of (meta)stability found in other high-pressure hydrides compared to experiments~\cite{heil2019superconductivity, kong2019superconductivity, errea2016quantum, einaga2016crystal}.

\section{Results}

\begin{figure*}[t]
	\centering
	\includegraphics[width=1\textwidth]{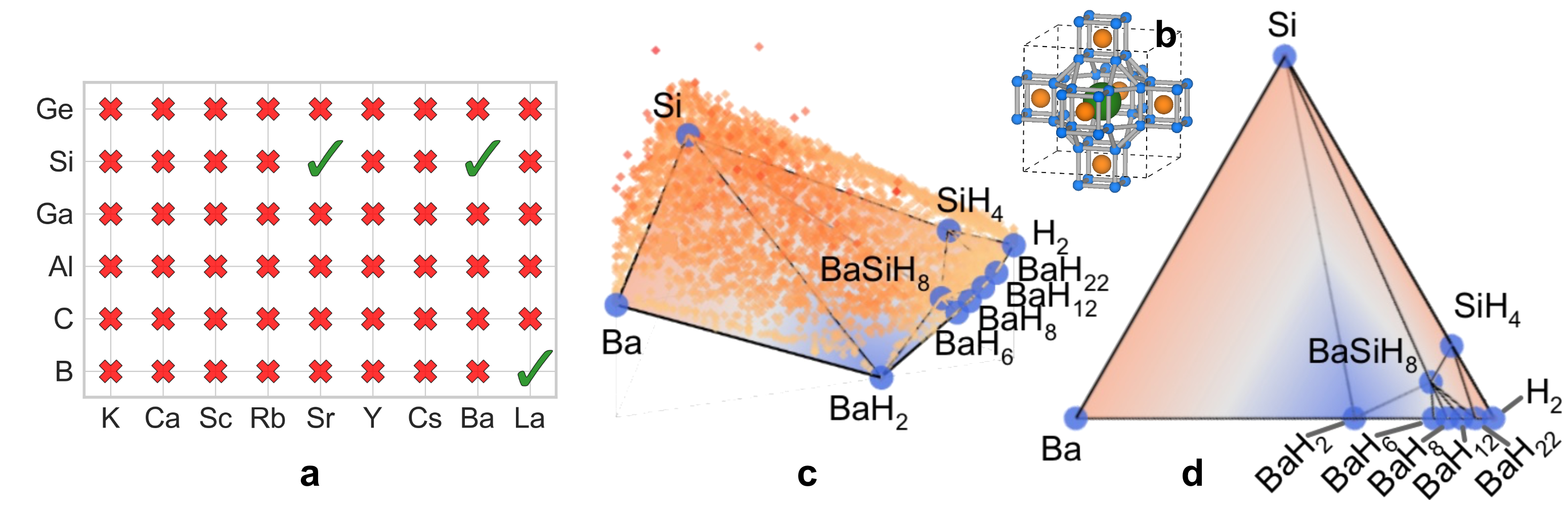}
	\caption{\textbf{Elemental composition and phase space search.} \textbf{a} Dynamical stability of the \XYH \ structure at \SI{50}{GPa} for various combinations of $X$ and $Y$ elements. 
	Dynamically stable and unstable compounds are indicated by red crosses	and green ticks, respectively. \textbf{b} Crystal structure of \fmm \ \XYH \ in the fcc conventional unit cell; H atoms are shown in blue, H-H bonds in grey, and the lighter (heavier) atom in orange (green). \textbf{c} Ternary convex hull for \BSH \ at \SI{100}{GPa} - side view; sampled structures are shown as orange dots. \textbf{d} Ternary convex hull for \BSH \ at \SI{100}{GPa} - top view. In \textbf{c} and \textbf{d} the color scale (from orange to blue) represents the \textit{depth} of the hull with respect to the elemental composition, from 0 to \SI{-0.7}{eV/atom}; stable compositions are shown as blue dots and are labeled.}
	\label{fig:convex_hull}
\end{figure*}

{\bf Initial screening:}
\BSH \ and \SSH \ were first identified through a high-throughput screening of possible  substitutions of La and B by neighbouring elements into the \fmm \ \XYH \ structure, as shown in Fig.~\ref{fig:convex_hull}\textbf{a}.
In this structure -- Fig.~\ref{fig:convex_hull}\textbf{b} -- hydrogen
atoms form rhombicuboctahedral cages around lanthanum; boron,
being much smaller, fills the six cubic voids surrounding the cages.
This structure permits a very efficient realization of a \textit{mechanical} precompression mechanism observed in SLC binary hydrides~\cite{liu2017potential,heil2019superconductivity}. To elaborate: in addition to the large central atom (La), also the smaller atom in the interstitials (B) exerts an additional pressure on the hydrogen sublattice, lowering the minimum stabilization pressure \pc~\cite{di2021labh}. 

We assumed that, considering different combinations of large ($X$) and small ($Y$) atoms, the superconducting properties of \LBH \ could be improved even further.
Aiming at identifying compounds with low critical pressures, we performed structural relaxations at \SI{50}{GPa} for all combinations of \mbox{$X = [\text{K, Rb,  Cs, Ca, Sr, Ba, Sc, Y, and La}]$} and \mbox{$Y = [\text{B, Al, Ga, C, Si, and Ge}]$} elements in the \fmm \ \XYH \ template, and evaluated the dynamical stability of the resulting compounds by calculating the harmonic phonon dispersions on a 4$\times$4$\times$4 grid in reciprocal space.

As shown in Supplementary~Figure~6 of the Supplementary Material (SM), the lattice constants of the relaxed structures exhibit an almost perfect linear dependence on the sum of the $X$ and $Y$ empirical atomic radii $R=R_X+R_Y$~\cite{slater_radii_table_1964}, with slopes determined by the total number of valence electrons $N_e=N_X+N_Y$.
Out of the 54 compounds investigated, only three compounds  with $N_e = 6$ -- green ticks in Fig.~\ref{fig:convex_hull}\textbf{a} -- are dynamically stable at \SI{50}{GPa}: Apart from the already reported \LBH~\cite{di2021labh}, we also found two silicon hydrides, \BSH \ and \SSH.
Both compounds exhibit a larger $R$ than \LBH, suggesting that mechanical precompression in both compounds will be more efficient, and hence their \pc \ may be lower. In particular, \BSH, where $R$ is considerably larger than in the other two compounds, looks very promising in this respect. Indeed, our harmonic phonon calculations yield  $p_\text{c}=5$ and \SI{30}{GPa} for \BSH\ and \SSH, respectively, lower than the $p_\text{c}=\SI{40}{GPa}$ in \LBH.

{\bf Convex hulls and phase diagrams:}
Having established that \BSH \ and \SSH \ are dynamically stable
in the \fmm \ structure close to ambient pressure, we determined the pressures at which the two compounds may spontaneously crystallize in the \fmm \ structure starting from appropriate precursors. 
To this end, we computed the full ternary convex hulls for the two ternary $X$-Si-H systems at ambient pressure and at \SI{100}{GPa}, employing \textit{ab-initio} variable-composition evolutionary crystal structure prediction methods as implemented in \textsc{Uspex}~\cite{glass_uspex_2006,oganov_how_2011}. 
To construct each hull, we sampled around 20000 structures and 5000 compositions, including the contribution of the zero-point energy (ZPE) from the ionic vibrations for structures close to the hull. Our results are in good agreement with previously reported structural searches on the Ba-H and Sr-H binary hydrides~\cite{chen_natcomm_2021_BaH, semenok_arxiv_2021_SrH22}.
Further computational details are given in Supplementary~Method~2 and Supplementary~Note~2 in the SM.
An example is illustrated in Fig.~\ref{fig:convex_hull}\textbf{c}-\textbf{d}, where we show the ternary convex hull of the Ba-Si-H system at \SI{100}{GPa}. (The analogous convex hull for the Sr-Si-H system at 0 and 100 GPa is shown in Supplementary~Figure~9 and 10 of the SM.) 

In the Ba-Si-H system, the \BSH \ composition is thermodynamically stable, i.e., it lies on the hull at \SI{100}{GPa} in a distorted $P1$ phase. The \fmm \ phase is slightly higher in enthalpy (\SI{32}{meV/atom}), but becomes enthalpically favourable above \SI{130}{GPa}. 
In the Sr-Si-H system at \SI{100}{GPa}, according to our calculations, \SSH \ should decompose into SrSiH$_6$ and SrSiH$_{12}$, which are the stable ternary compositions along with Sr$_2$SiH$_{10}$. The corresponding structures are shown in Supplementary~Figure~12 of the SM. However, the \fmm \ phase lies only \SI{54}{meV/atom} above the convex hull. From a comparison of the formation enthalpies, we predict that it should become stable at \SI{174}{GPa}.

{\bf Anharmonic lattice dynamics:}
\begin{figure*}[t]
	\centering
	\includegraphics[width=1\textwidth]{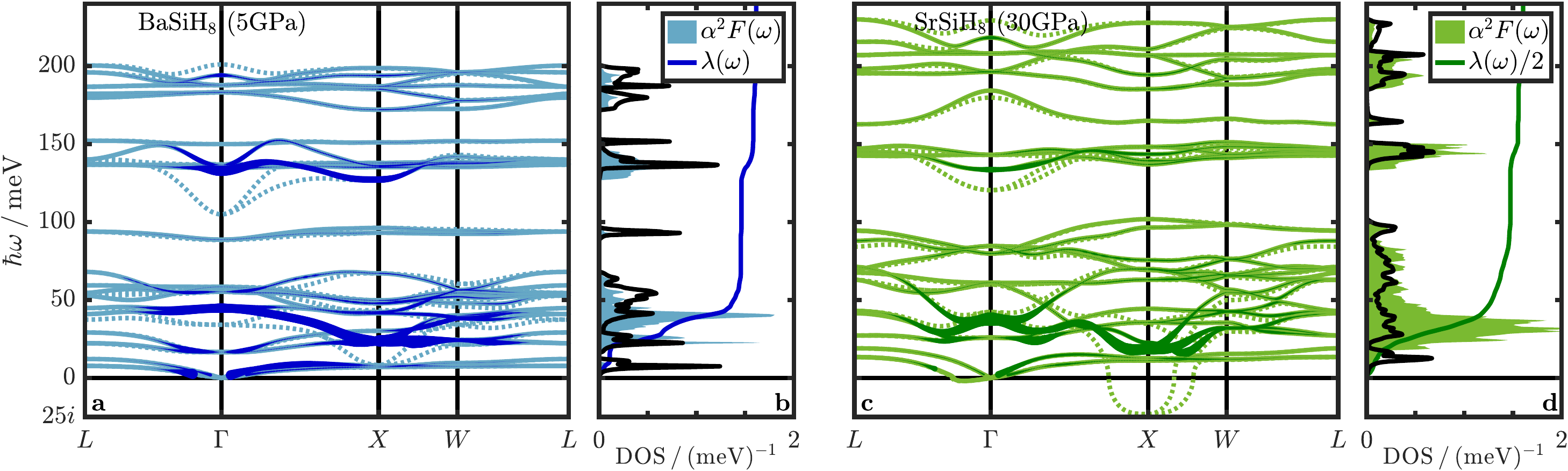}
	\caption{\textbf{Vibrational structure and electron-phonon properties.} \textbf{a} Phonon dispersion of \BSH \ at \SI{5}{GPa};  the thickness of the dark blue lines is proportional to the mode- and wave-vector resolved coupling constant $\lambda_{\mathbf{q},\nu}$. Solid (dotted) lines correspond to anharmonic (harmonic) results. \textbf{b} Phonon DOS shown as solid black line, $\alpha^2F(\omega)$ as filled curve, and $\lambda(\omega)$ as solid blue line. Panels \textbf{c} and \textbf{d} as \textbf{a} and \textbf{b} but for \SSH \ at \SI{30}{GPa}.
	}
	\label{fig:ph_bands}
\end{figure*}
\begin{figure*}[t]
	\centering
	\includegraphics[width=1\textwidth]{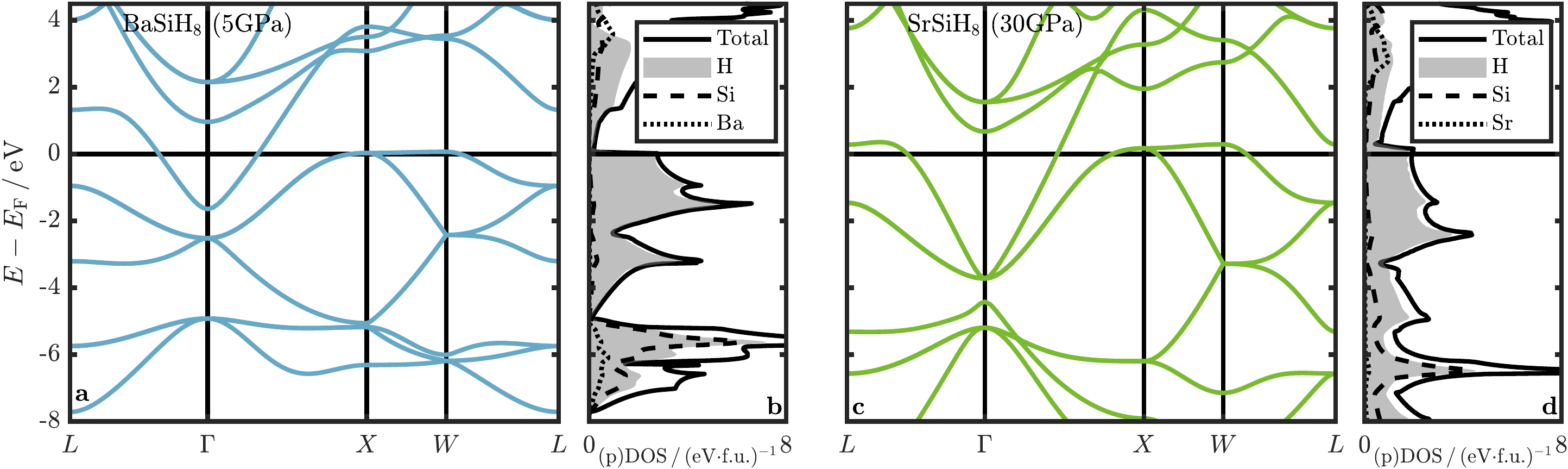}
	\caption{\textbf{Electronic band structure and density of states.} \textbf{a} Electronic band structure of \BSH \ at \SI{5}{GPa} and \textbf{b} corresponding total DOS (solid black) and projected H DOS (filled blue). Panels \textbf{c} and \textbf{d} as \textbf{a} and \textbf{b} but for \SSH \ at \SI{30}{GPa}.
	}
	\label{fig:el_bands}
\end{figure*}
Preliminary phonon calculations in the harmonic approximation indicated that both \BSH \ and \SSH \ experience remarkable phonon softening at lower pressures. More specifically, we find soft phonon modes, mainly at $\Gamma$ ($T_{2g}$, $E_g$, and $A_{1g}$) and $X$ ($E_{g}$, and additionally $A_{2g}$ in \SSH), as shown in Fig.~\ref{fig:ph_bands}. 
Eventually, imaginary frequencies appear for the $E_g$ mode at $X$ below \SI{5}{GPa} for \BSH \ and \SI{40}{GPa} for \SSH.  (A full set of phonon dispersions at all pressures considered is provided in the SM.)

Soft-mode behaviour, associated with strong anharmonicity, 
has been reported for many hydrogen-rich materials~\cite{errea2016quantum,errea2020quantum,borinaga2016anharmonic,heil_influence_2015,heil2019superconductivity}. 
Anharmonic lattice effects have been shown to crucially affect the range of dynamical stability, phonon frequencies and eigenvectors of superhydrides, and their inclusion is essential to obtain accurate estimates of these quantities.

In order to account for anharmonic effects on the phonon dispersions 
of \BSH \ and \SSH, we evaluated the adiabatic potential energy surface (APES) for every soft mode of a 2$\times$2$\times$2 wave-vector grid, which includes explicitly the special points $\Gamma$, $L$, and $X$, and solved the resulting Schr\"odinger equations. (More information can be found in Supplementary Method 4 of the SM, and Ref.~\cite{heil_origin_2017}.)

The harmonic and the anharmonic phonon dispersions for \BSH \ and \SSH \ close to their critical pressures of stability are shown as dashed and full lines in Fig.~\ref{fig:ph_bands}.   Anharmonicity causes a considerable hardening of the $T_{2g}$ (at $\Gamma$) and $E_g$ (at $\Gamma$ and $X$) modes, leading to a decrease of the critical pressure of dynamical stability \pc: Taking this hardening into account, \SSH \ is stable down to a pressure of \SI{27}{GPa} and, even more excitingly, \BSH \ down to \SI{3}{GPa}.

{\bf Electronic structure and superconductivity:}
Having determined that both \BSH \ and \SSH \ remain dynamically stable close to ambient pressure, we are left with the question whether at these relatively low pressures they can still be considered superhydrides. A first positive indication comes from the analysis of their electronic structure.

In Fig.~\ref{fig:el_bands} we show the electronic dispersions and DOS for \BSH \ (left) and \SSH \ (right), at 5 and \SI{30}{GPa}, respectively.
Despite a scaling of the total bandwidth due to the different pressures, the band structures are qualitatively very similar. The Fermi level is located just above a large DOS shoulder; in this region and up to \SI{-5.5}{eV} for \BSH \ (\SI{-3.5}{meV} for \SSH),
the bands are of purely hydrogen character.
This is an important prerequisite for high-\Tc \ conventional
superconductivity in hydrides, since it can imply a strong electron-phonon ($ep$) coupling between electronic states to the light hydrogen sublattice~\cite{heil_absence_2018}. 

Similar conclusions can be inferred from the electron localization function (ELF) of the two compounds (see Supplementary~Figure~15 in the SM), where, in line with what we observed in \LBH~\cite{di2021labh}, we find an increased charge localization in the vicinity of the H and Si atoms and between H-H, but not between Sr/Ba and H or Si and H.
This supports the idea that neither Sr/Ba, nor Si form bonds with H, and thus act on the H sublattice essentially as \textit{mechanical} spacer, as in binary sodalite-like clathrate hydrides~\cite{heil2019superconductivity,errea2020quantum,flores2020perspective,belli2021strong}.

\begin{figure}[t]
	\centering
 	\includegraphics[width=1\columnwidth]{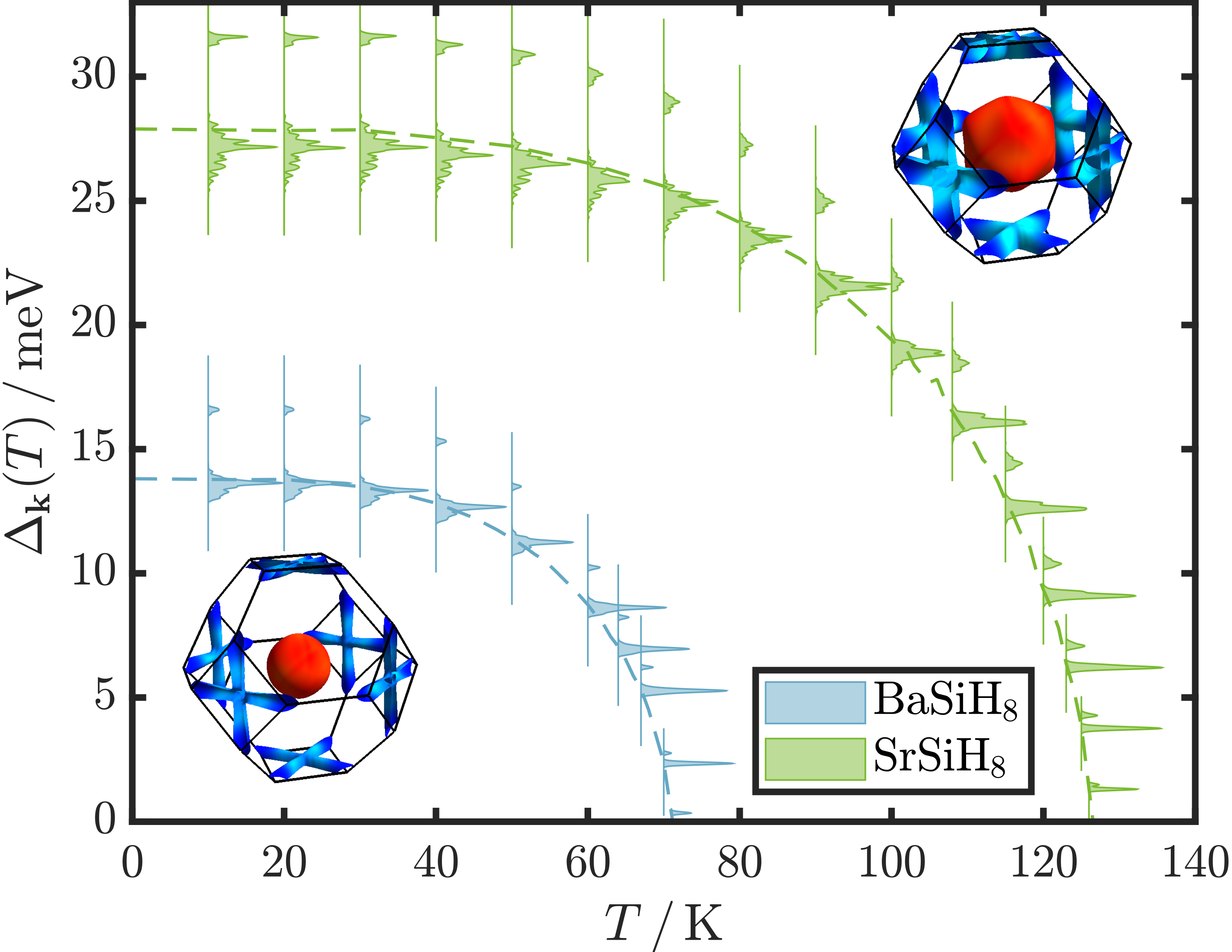}
	\caption{\textbf{Superconducting gap as a function of temperature.} $\Delta(T)$ for \BSH \ (blue) at \SI{5}{GPa} and for \SSH \ (green) at \SI{30}{GPa} calculated using the fully anisotropic ME theory; dashed lines indicate the mean value as a guide to the eye. The insets in the top right and bottom left corner show the Fermi surface  distribution of $\Delta_\mathbf{k}$ at \SI{10}{K} for \BSH \ and \SSH \, respectively; blue/red correspond to  $\text{min}(\Delta)$/$\text{max}(\Delta)$.
	}
	\label{fig:Delta_vs_T}
\end{figure}

The Fermi surface topology is the same in the two compounds:
a large, spheric-like electron pocket is centered around the Brillouin zone center, while a more complicated hole-like network extends around the faces of the Brillouin zone, enclosing the $X$ and $W$ points (see the insets of Fig.\ref{fig:Delta_vs_T}).

These qualitative electronic structure arguments are confirmed by actual calculations of the superconducting properties.
Fig.~\ref{fig:Delta_vs_T} shows the energy distribution of the superconducting gap $\Delta$ as a function of temperature $T$ for \BSH \ at \SI{5}{GPa} pressure and for \SSH \ at \SI{30}{GPa}, obtained by solving the anisotropic Migdal-Eliashberg (ME) equations on a 30$\times$30$\times$30 $\mathbf{k}-$ and $\mathbf{q}$-grid using the anharmonically corrected phonon dispersions with the \textsc{Epw} code~\cite{margine_anisotropic_2013,ponce_epw_2016}. 
(The superconducting properties at all other considered pressures and further computational details are provided in the SM.)
We observe two distinct superconducting gaps: The inset of the figure shows that large $\Delta$ values correspond to the $\Gamma$-centered, spherical electron pocket, while lower values occur on the hole-like tubular network around the $X$ and $W$ points. The superconducting critical temperatures are predicted to be 71 and \SI{126}{K} for \BSH \ and \SSH, respectively. 

\begin{figure}[t]
	\centering
	\includegraphics[width=1\columnwidth]{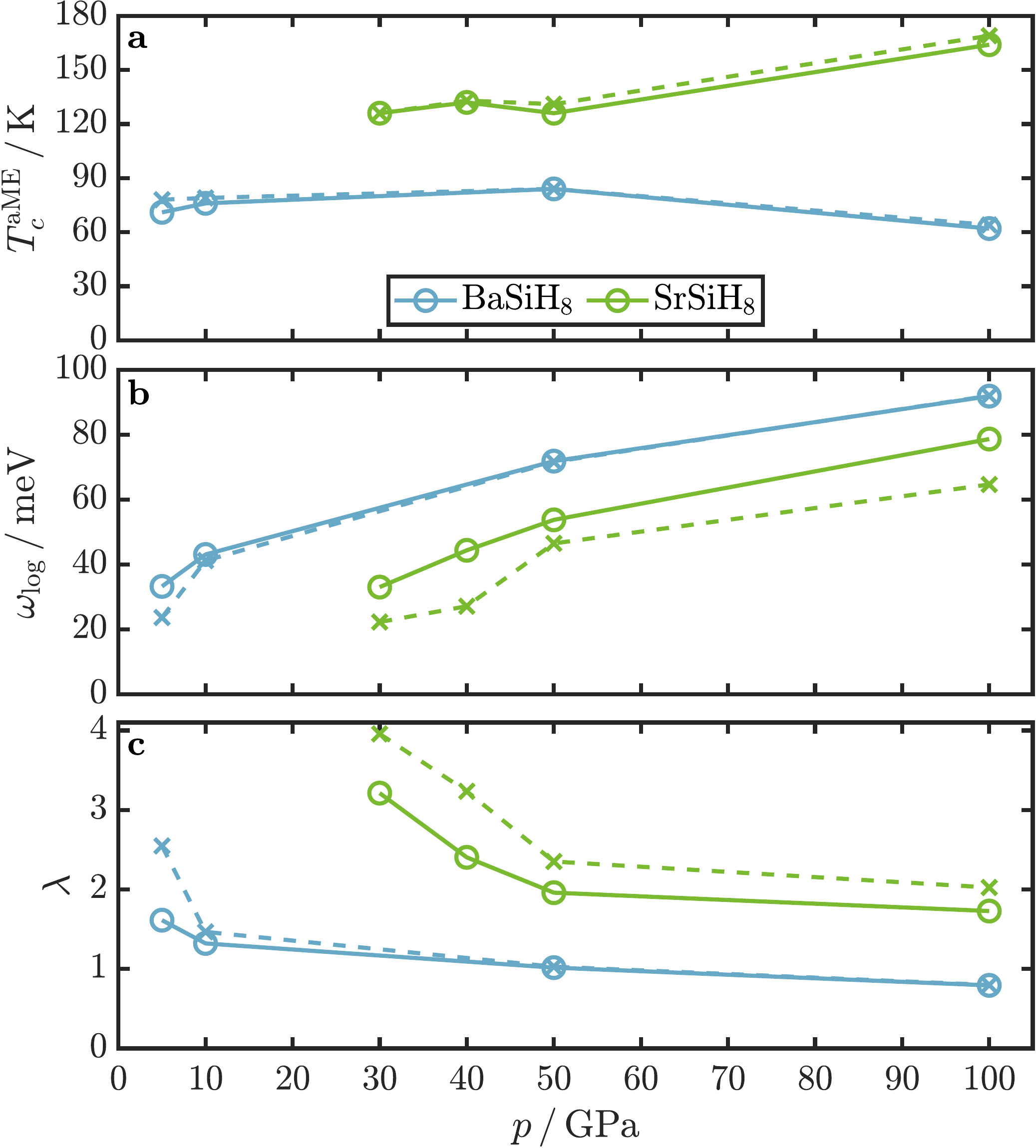}
	\caption{\textbf{Superconducting properties as a function of pressure.} \textbf{a} \Tc \ from anisotropic ME theory, \textbf{b} $\omega_\text{log}$, and \textbf{c} $\lambda$ (bottom panel) as functions of pressure for \BSH \ (blue) and \SSH \ (green). Circles and solid lines (crosses and dashed lines) indicate anharmonically corrected (harmonic) calculations. The contributions to $\alpha^2 F(\omega)$ from the $E_g$ mode at $X$ are set to zero in the harmonic calculations for \SSH \ at 30 and \SI{40}{\giga\pascal}, as the corresponding phonon frequencies are imaginary.
	}
	\label{fig:SC_vs_pressure}
\end{figure}

Further details on the origin of the remarkable \Tc's of \BSH \ and \SSH \ can be obtained from an analysis of the distribution of their $ep$ coupling over phonon branches.
In panel \textbf{b} of Fig.~\ref{fig:ph_bands}, the mode and wave-vector resolved $ep$ coupling $\lambda_{\mathbf{q},\nu}$ are overlayed onto the phonon dispersions as fat bands; panel \textbf{d} of the same figure shows the Eliashberg spectral function $\alpha^2F(\omega)$, and the total frequency-dependent $ep$ coupling parameter $\lambda (\omega)$.

\begin{figure*}[t]
	\centering
	\includegraphics[width=1\textwidth]{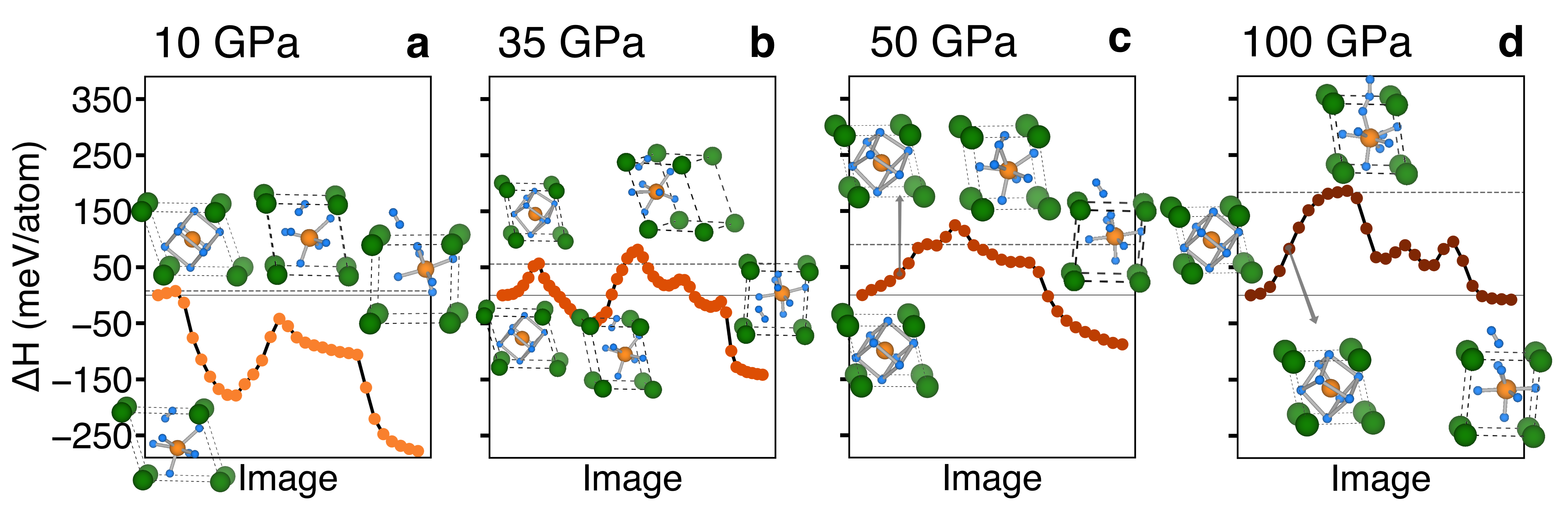}
	\caption{\textbf{VCNEB calculations for \BSH \ at different pressures.} Minimum energy path (MEP) for \BSH \ at various pressures, calculated using the VCNEB method. Each  point represents an individual crystal structure (image) along the path. Particularly relevant intermediate states are shown.}
	\label{fig:metastability}
\end{figure*}

The figure clearly shows that the largest contributions to the total coupling come from low-energy modes; a substantial fraction is associated to soft phonon modes around $\Gamma$ and $X$. In the case of \BSH, for example, we estimate that the phonons related to $\Gamma-T_{2g}$ contribute roughly \SI{40}{\%} to the total $\lambda$ and around $X-E_g$ about \SI{35}{\%}. These modes are of purely hydrogen character and their patterns can be more easily visualized in terms of the H cubes around Si: The $T_{2g}$ mode at $\Gamma$ is a centrosymmetric stretching and squeezing along one of the space diagonals of the H cube, while the $E_g$ mode at $X$ is a non-deforming rotation around a face normal of the H cube, with alternating phase in two neighbouring cells. 

Fig.~\ref{fig:SC_vs_pressure} shows the evolution of the superconducting properties of \BSH \ and \SSH \ as a function of pressure from the lowest dynamical stability pressure up to \SI{100}{GPa}; open circles and solid lines correspond to anharmonic results, while dashed lines refer to harmonic values.
Interestingly, \Tc \ is approximately constant with pressure -- panel \textbf{a}; this is true even close to \pc, where one would expect a strong deviation due to anharmonicity.
An inspection of panels \textbf{b} and \textbf{c} shows that this happens because of the compensating effect of phonon hardening on $\omega_\text{log}$ and $\lambda$. The effect is more marked in \SSH \ and in \BSH \ close to the instability point, where harmonic and anharmonic dispersions diverge the most. For \SSH \ at \SI{30}{\giga\pascal}, we find an anharmonic (harmonic) $\lambda$ of 3.21 (3.96) and an $\omega_\text{log}$ of \SI{33.1}{\milli\electronvolt} (\SI{22.3}{\milli\electronvolt}), and for \BSH \ at \SI{5}{\giga\pascal}, we find a $\lambda$ of 1.61 (2.55) and an $\omega_\text{log}$ of \SI{33.3}{\milli\electronvolt} (\SI{23.7}{\milli\electronvolt}).

{\bf Metastability:}
In all theoretical studies of high-pressure hydrides performed so far, the range of metastability of high-pressure phases has been assumed to coincide with the range of dynamical stability, eventually including quantum  corrections to lattice dynamics.
However, a comparison of these predictions 
and available experimental data suggests that theoretical estimates of
the critical pressure are systematically lower than experimental values~\cite{heil2019superconductivity, kong2019superconductivity, errea2016quantum, einaga2016crystal}.

Indeed, dynamical stability is only a prerequisite for thermodynamic
metastability. The latter is determined by the existence of an energy (enthalpy) barrier protecting a metastable phase from decomposition into other phases~\cite{ceder_metastability_2016}. Attempts to quantitatively estimate the barrier height are extremely rare.
For \fmm \ \BSH, where the dynamical instability pressure \pc \ is extremely close to ambient pressure, this issue is obviously crucial, since the presence of a sufficiently high barrier could be considered the definitive proof of synthesizability. 

In this work, we estimated the barrier height using the VCNEB as implemented in the \textsc{Uspex} code~\cite{oganov_2013_VCNEB}.
In this method, a number of intermediate structures (\textit{images}) are created between the metastable structure of interest and the ground-state structure at a given pressure; elastic forces between each image are added to the `physical' forces, and the whole chain of images thus created is finally relaxed to obtain the energy/enthalpy profile of the transition. 

We ran VCNEB simulations for \BSH\ at six different pressures: 10, 25, 35, 50, 100, and \SI{200}{GPa}. As end-members for the VCNEB path, we chose the \fmm \ phase and a $P1$ phase, identified as the ground-state structure at \SI{10}{GPa} through a fixed-composition structural search. 
Due to its large unit cell and low symmetry, it can be assumed that the $P1$ phase, relaxed at the different pressures, approximates quite well the true ground-state of the system; indeed, at all pressures the $P1$ \BSH \ phase lies only a few meV above the hull. In practice, analyzing the VCNEB images in Fig.~\ref{fig:metastability}, we observe that the transition from the \fmm \ to the $P1$ phase corresponds to the decomposition of \BSH \ into BaSiH$_6$ + H$_2$, with the expulsion of hydrogen in molecular form.

The potential barrier, shown in Supplementary~Figure~3, decreases  with pressure from \SI{153}{meV/atom} at \SI{100}{GPa} to \SI{57}{meV/atom} at \SI{35}{GPa}; a sharp transition is visible at \SI{25}{GPa}, where
the kinetic barrier abruptly drops to \SI{9}{meV/atom}. Although a small barrier survives down to \SI{10}{GPa}, this sharp decrease is the signature of an impending \textit{kinetic} instability, i.e., the metastable state will be short-lived and most likely will not be observed in experiments.

Combining the convex hull results in Fig.~\ref{fig:convex_hull}  with the VCNEB analysis, we can argue with confidence that an \fmm \ \BSH \ phase could be synthesized above \SI{100}{GPa}, and retained down to $\sim$\SI{30}{GPa}, where a clearly visible enthalpy barrier exists. At lower pressures, metastable \fmm \ \BSH \ will decompose, even though (anharmonic) lattice dynamics calculations predict it to be stable.
Hence, \textit{kinetic} stability poses a stricter bound for synthesizability than  \textit{dynamical} stability.

In addition to the promising theoretical synthesizability, both materials appear to be convenient in the experimental setting. As the ambient-pressure convex hulls in Figs.~S7 and S9 reveal, both systems feature known stable orthorhombic binary monosilicides, namely BaSi and SrSi~\cite{pani_2008_BaSi,palenzona_2004_SrSi}, which serve as adequate starting materials having the target composition of Ba/Sr and Si. The monosilicide can be loaded in a diamond anvil cell together with the common solid hydrogen storage medium ammonia borane (H$_3$NBH$_3$) which releases the hydrogen upon heating and forms the refractory compound boron nitride (BN).

{\bf Discussion:}
In summary, our study  shows that indeed the superconducting properties of the \XYH \ template can be optimized by a suitable choice of the $X$ and $Y$ elements.
The two silicides identified in this work represent an improvement compared to \LBH: for \SSH \ we predict a dynamical stability pressure \pc \ of \SI{27}{GPa}, with a \Tc \ of \SI{127}{K}.
Using the figure of merit $S$ introduced by the authors of Ref.~\cite{pickard2020superconducting}, this means passing from 1.3 in H$_3$S and LaH$_{10}$ to 2.2 in \LBH \ to 2.7 in \SSH.
Even more remarkably, \BSH \ is predicted to be dynamically stable down to pressures of \SI{3}{GPa}, with a critical temperature of \SI{71}{K}, which is substantially higher than all established~\cite{nagamatsu_mgb2_2001} and claimed~\cite{bhaumik_qcarbon_2017} experimental \Tc \ records for conventional superconductors at ambient pressure.

VCNEB calculations demonstrate that the shape of the potential energy landscape of \BSH \ is favorable for its synthesis in a wide range of pressures. In contrast to the binary La-B system, the Ba-Si system features also the ideal starting material for the synthesis, namely the stable monosilicide BaSi, and a possible path involves synthesis at high pressure ($p > \SI{130}{GPa}$) and/or laser heating, and rapid quenching of the resulting phase to lower pressure, i.e., down to $\sim$\SI{30}{GPa}, where the metastable \fmm \ crystal structure is protected by a sizable potential barrier.
This defines a {\em kinetic} threshold pressure, $p_\text{kin}$,  which is substantially higher than \pc.

By defining a concrete method to estimate the synthesizability of a proposed structure, our study sets a new standard for the \textit{ab-initio} design of new superconductors at high pressures, based on the more rigorous concept of \textit{kinetic} stability,
rather than dynamical stability.
The existence of  a distinct {\em kinetic} stability criterion may also be invoked to explain why many long-standing predictions of high-\Tc \ superconductors have not been realized experimentally~\cite{libc2002,graphane2010,saha2020high}.

We strongly believe that the proposed method represents a major step forward  towards achieving  high-\Tc \ conventional superconductivity at room pressure.

\section{Methods}
\subsection{Crystal structure prediction}
Crystal structure prediction runs were carried out using evolutionary algorithms as implemented in the \mbox{\textsc{Uspex}} package~\cite{glass_uspex_2006,oganov_how_2011,lyakhov_new_2013}; the underlying total energy and structural relaxation calculations were performed using plane-waves and pseudopotentials as implemented in the \mbox{Vienna} \textit{ab-initio} Simulation Package \textsc{Vasp}~\cite{kresse_efficiency_1996}. 
Further computational details are provided in Supplementary~Method~2 of the SM. 
\subsection{Electronic and vibrational properties}
All density-functional theory (DFT) and  density-functional perturbation theory (DFPT) calculations of electronic and vibrational properties were carried out using the plane-wave pseudoptential code \textsc{Quantum Espresso}~\cite{giannozzi_advanced_2017}, scalar-relativistic optimized norm-conserving Vanderbilt pseudopotentials (ONCV)~\cite{hamann_optimized_2013}, and the PBE-GGA exchange and correlation functional~\cite{perdew_generalized_1996}. Computational details are provided in Supplementary~Method~1 of the SM.
\subsection{Phase transition paths}
The phase transition path between the $P1$ and \fmm-\BSH \ was evaluated using the VCNEB method as implemented in \textsc{Uspex}~\cite{glass_uspex_2006, oganov_2013_VCNEB}, using variable elastic constants and a variable number of images between the endpoints. The energy and forces were calculated using \textsc{Vasp}. Further computational details are provided in Supplementary~Method~3 of the SM.
\subsection{Migdal-Eliashberg theory}
The interpolation of the $ep$ matrix elements onto dense wave-vector grids and the subsequent self-consistent solution of the fully anisotropic Migdal-Eliashberg equations were done with \textsc{Epw}~\cite{ponce_epw_2016}. Based on our previous work on \LBH,  where we calculated the Morel-Anderson pseudopotential $\mu^*$ from first principles using $GW$ and found consistent values for $\mu^*$ of about 0.1~\cite{heil2019superconductivity,heil_absence_2018,di2021labh}, we chose the same value in the current work. Further computational details are provided in Supplementary~Method~5 of the SM. 
\subsection{Phonon anharmonicity}
Anharmonic corrections to phonon frequencies were obtained by explicitly solving the Schr\"odinger equation for the APES of every soft mode on a 2$\times$2$\times$2 wave-vector grid. By calculating and solving 2D APES, we also checked that phonon-phonon interactions for various modes and wave vectors can be neglected in a good first approximation. 
The interpolation of the real-space force constants obtained on the 2$\times$2$\times$2 $\mathbf{q}$-grid was performed using the corresponding harmonic support DFPT solution as implemented in the \textsc{Cell}\textsc{Constructor} package~\cite{monacelli2021stochastic}.
Further computational details are provided in Supplementary~Method~4 of the SM.

\section{Data availability}
The authors confirm that the data supporting the findings of this study are available within the article and its supplementary materials. Further information is available upon request.

\section{Acknowledgments}
We thank Dmitrii Semenok (Skolkovo Institute of Science and Technology) for pointing out the experimental convenience of the two proposed materials.

This research was funded by the Austrian Science Fund (FWF) P~30269-N36 and P~32144-N36. For the purpose of open access, the authors have applied a CC~BY public copyright license to any Author Accepted Manuscript version arising from this submission. This work was supported by the dCluster of the Graz University of Technology and the VSC-4 of the Vienna University of Technology. L.B. acknowledges support from Fondo Ateneo-Sapienza 2017-2019. S.D.C. acknowledges computational resources from CINECA, proj. IsC90-HTS-TECH\_C

\section{Author contributions}
R.L. and S.D.C. performed the calculations, and L.B. and C.H. conceived and supervised the project. All authors contributed to the discussion of the results and participated in preparing the manuscript.

\section{Competing interests}
The authors declare no competing financial or non-financial interests.

%

\clearpage
\onecolumngrid
\newpage
\printfigures

\end{document}